\documentstyle[epsf,12pt]{article}

\begin{document}

\hoffset = -1truecm
\voffset = -2truecm
\baselineskip = 10 mm

\title{\bf Antishadowing effects in the unitarized BFKL equation}

\author{
{\bf Jianhong Ruan}, {\bf Zhenqi Shen}, {\bf Jifeng Yang} and {\bf
Wei Zhu\footnote{Corresponding author, E-mail: weizhu
@mail.ecnu.edu.cn}}\\
\normalsize Department of Physics, East China Normal University,
Shanghai 200062, P.R. China \\
}

\date{}

\newpage

\maketitle

\vskip 3truecm

\begin{abstract}

A unitarized BFKL equation incorporating shadowing and
antishadowing corrections of the gluon recombination is proposed.
This equation reduces to the Balitsky-Kovchegov evolution equation
near the saturation limit. We find that the antishadowing effects
have a sizeable influence on the gluon distribution function in
the preasymptotic regime.

\end{abstract}


$keywords$:  antishadowing; evolution equation; QCD recombination
process

\newpage
\begin{center}
\section{Introduction}
\end{center}

   The growth of cross sections with gluon splitting according to the DGLAP [1]
and BFKL [2] evolution equations would violate the unitarity.
Therefore, the corrections of the higher order QCD, which shadow
the growth of parton densities and lead to an eventual saturation
of parton densities, become a focus of intensive study in recent
years. In this respect, the GLR-MQ (by Gribov, Levin Ryskin in [3]
and Mueller and Qiu in [4]) and BK (by Balitsky and Kovchegov in
[5]) equations are broadly regarded as two solvable models to
compute the shadowing corrections to the DGLAP and BFKL equations,
respectively.

       One of the key problems in restoring unitarity is the origin of the negative
corrections. In the viewpoint of elementary QCD interaction, the
suppression of the gluon splitting comes from its inverse
process-the gluon recombination. The negative screening effects in
the recombination processes originally occur in the interferant
cut-diagrams of the recombination amplitudes [3,6]. For computing
the contributions from the interference processes, the AGK cutting
rules [7] were used in the derivation of the GLR-MQ equation.

       On the other hand, one of us (Zhu) disputed the above
mentioned applications of the AGK cutting rules in the GLR-MQ
equation, and used the TOPT-cutting rules based on the time
ordered perturbation theory (TOPT) instead of the AGK cutting
rules to expose the relations among various cut diagrams in a
recombination process [6]. Thus, we can completely compute the
contributions of the gluon recombination processes. Consequently,
a modified DGLAP equation was proposed [6,8]. A remarkable
property of this equation is that the positive antishadowing and
negative shadowing components in the nonlinear evolution equation
are naturally separated. As a result, the corrections from gluon
recombination at small $x$ depend not only on the size of gluon
density at this value of $x$, but also on the shape of gluon
density in the region $[x/2,x]$. Thus, the shadowing effects in
the evolution process will be weakened by the antishadowing
effects if the gluon distribution has a steeper form.

       The antishadowing effects always coexist with the shadowing
effects in the QCD recombination processes as a general conclusion
of the momentum conservation [9]. Therefore, similar antishadowing
effects should exist in any unitarized BFKL equation. In this work
we try to find the antishadowing effects in the unitarized BFKL
equation. The total momentum carried by gluons is not changed as a
result of the gluon fusion but simply redistributed to different
$x$ regions. The depleted momentum due to the fusion is heaped up
step by step toward a larger $x$ direction, which leads to an
enhanced density of higher momentum gluons (i.e. the antishadowing
effects). We shall show that the antishadowing effects have a
sizeable influence on the gluon distribution function in the
preasymptotic regime.

    The paper is organized as follows.  In Section 2 we derive a
unitarized BFKL equation with the gluon recombination, which
contains the contributions of the shadowing and antishadowing
effects. The numerical analysis of our equation are presented in
Section 3. In Section 4 we show that our equation reduces to the BK
equation near the saturation limit.  We also compare our equation
with a modified BK equation in this Section.

\newpage
\begin{center}
\section{The evolution equation incorporating shadowing and antishadowing effects}
\end{center}

     We develop a unified framework to construct the evolution equations
for both the integrated and unintegrated gluon distribution
functions. We begin from a deep inelastic scattering process,
where the unintegrated gluon distribution is measured. In the
$k_T$-factorization scheme, the cross section is decomposed into

$$d\sigma(probe^* P\rightarrow k'X)$$
$$=f(x_1,\underline{k}_1^2)\otimes{\cal
K}\left(\frac{\underline{k}^2}{\underline{k}_1^2},\frac{x}{x_1},\alpha_s\right)\otimes
d\sigma(probe^*k\rightarrow k')$$
$$\equiv \Delta f(x,\underline{k}^2)\otimes d\sigma(probe^*k\rightarrow
k'), \eqno(1)$$ which contains the perturbative evolution kernel
${\cal K}$, the nonperturbative unintegrated gluon distribution
function $f$ and the $probe^*$-parton cross section
$d\sigma(probe^*k\rightarrow k')$. For simplicity, we take the
fixed QCD coupling in this work. According to the scale-invariant
parton picture of the renormalization group [10], we regard
$\Delta f(x,\underline{k}^2)$ as the increment of the distribution
$f(x_1,\underline{k}_1^2)$ when it evolves from $(x_1,\underline
{k}_1^2)$ to $(x,\underline {k}^2)$. Thus, the connection between
the two functions $f(x_1,\underline{k}_1^2)$ and
$f(x,\underline{k}^2)$ via Eq.(1) is

$$f(x,\underline{k}^2)=f(x_1,\underline{k}^2_1)+\Delta
f(x,\underline{k}^2)$$
$$=f(x_1,\underline{k}^2_1)+\int^{\underline{k}^2}_{\underline{k}^2_{1{min}}}\frac{d\underline{k}^2_1}{\underline{k}^2_1}\int^1_{x}\frac{dx_1}{x_1}
{\cal
K}\left(\frac{\underline{k}^2}{\underline{k}_1^2},\frac{x}{x_1},\alpha_s\right)
f(x_1,\underline{k}^2_1), \eqno(2) $$

        The relation of the unintegrated gluon distribution with the integrated
gluon distribution is

$$G(x,Q^2)\equiv
xg(x,Q^2)=\int^{Q^2}_{\underline{k}^2_{min}}\frac{d\underline{k}^2}{\underline{k}^2}
xf(x,\underline{k}^2)\equiv\int^{Q^2}_{\underline{k}^2_{min}}\frac{d\underline{k}^2}{\underline{k}^2}
F(x,\underline{k}^2), \eqno(3)$$ or

$$f(x,\underline{k}^2)=Q^2\frac{\partial g(x,Q^2)}{\partial
Q^2}\left |_{Q^2=\underline{k}^2}\right .. \eqno(4)$$

  In the evolution along the transverse momentum (or along the longitudinal momentum),
we differentiate Eq.(2) with respect to $\underline{k}^2$ (or to
$x$) and get

$$\frac{\partial f(x,\underline{k}^2)}{\partial \underline{k}^2}$$
$$=\left .\int^1_{x}\frac{dx_1}{x_1}\frac{1}{\underline{k}^2_1}
{\cal
K}\left(\frac{\underline{k}^2}{\underline{k}_1^2},\frac{x}{x_1},\alpha_s\right)
f(x_1,\underline{k}_1^2) \right |
_{\underline{k}^2_1=\underline{k}^2}$$
$$+\int^{\underline{k}^2}_{\underline{k}^2_{1{min}}}
\frac{d\underline{k}_1^2}{\underline{k}_1^2}
\int^1_{x}\frac{dx_1}{x_1}\frac{\partial {\cal
K}\left(\frac{\underline{k}^2}{\underline{k}_1^2},\frac{x}{x_1},\alpha_s\right)}{\partial
\underline{k}^2} f(x_1,\underline{k}^2_1),  \eqno(5)$$ or

$$-\frac{\partial f(x,\underline{k}^2)}{\partial x}$$
$$=\left.\int^{\underline{k}^2}_{\underline{k}^2_{1{min}}}
\frac{d\underline{k}_1^2}{\underline{k}_1^2}\frac{1}{x_1} {\cal
K}\left(\frac{\underline{k}^2}{\underline{k}_1^2},\frac{x}{x_1},\alpha_s\right)
f(x_1,\underline{k}_1^2) \right | _{x_1=x}$$
$$-\int^{\underline{k}^2}_{\underline{k}^2_{1{min}}}
\frac{d\underline{k}_1^2}{\underline{k}_1^2}
\int^1_{x}\frac{dx_1}{x_1}\frac{\partial {\cal
K}\left(\frac{\underline{k}^2}{\underline{k}_1^2},\frac{x}{x_1},\alpha_s\right)}{\partial
x} f(x_1,\underline{k}^2_1), \eqno(6)$$ respectively. Generally,
the resummation solution is hard to obtain from these two
equations. However, at the $LL(\underline{k}^2)A$ (or at the
$LL(1/x)A$) the evolution kernel $\cal K$ in Eq.(5) (or in Eq.(6))
is only the function of the longitudinal variable (or of the
transverse variables) and the second term on the right-hand side
of Eq.(5) (or of Eq.(6)) vanishes. In this case, the resummation
becomes possible.  For example, using Eq.(3) we write

$$\Delta g(x,Q^2)\equiv\int^{Q^2}_{\underline{k}^2_{min}}\frac{d\underline{k}^2}{\underline{k}^2}
\Delta f(x,\underline{k}^2)$$
$$=\int^{Q^2}_{\underline{k}^2_{min}}\frac{d\underline{k}^2}{\underline{k}^2}
\int^{\underline{k}^2}_{\underline{k}^2_{1min}}\frac{d\underline{k}^2_1}{\underline{k}^2_1}
\int_{x}^1\frac{dx_1}{x_1}{\cal
K}_{DGLAP}\left(\frac{x}{x_1},\alpha_s\right)f(x_1,\underline{k}^2_1)$$
$$=\int^{Q^2}_{\underline{k}^2_{min}}\frac{d\underline{k}^2}{\underline{k}^2}
\int_{x}^1\frac{dx_1}{x_1} {\cal
K}_{DGLAP}\left(\frac{x}{x_1},\alpha_s\right)g(x_1,\underline{k}^2),
\eqno(7)$$ and

$$g(x,Q^2)=g(x_1,\underline{k}^2)+\Delta g(x,Q^2). \eqno(8)$$
On the other hand, at $DLL(Q^2)A$ we have

$${\cal K}_{DGLAP}\frac{dx_1}{x_1}
=\frac{\alpha_sN_c}{\pi} \frac{dx_1}{x}. \eqno(9)$$ Thus, from
Eqs.(5) and (7) one can get the DGLAP equation at small $x$ for
the gluon distribution

$$Q^2\frac{\partial g(x,Q^2)}{\partial Q^2} =\int_{x}^1\frac{dx_1}{x_1}
{\cal K}_{DGLAP}\left(\frac{x}{x_1},\alpha_s\right)g(x_1,Q^2)$$
$$=\frac{\alpha_sN_c}{\pi}\int_{x}^1\frac{dx_1}{x_1}
\frac{x_1}{x} g(x_1,Q^2). \eqno(10)$$

     Now let us consider the corrections of the gluon recombination to the DGLAP equation.
The one step evolution containing the gluon recombination is
illustrated in Fig.1(a), where the initial gluons from the two
legs are resummed using the DGLAP equation. Four different kinds
of the evolution kernels for ${\cal {K}}_{MD-DGLAP}$ are listed in
Fig.2. Similar to Eqs.(7) and (8) we derive the modified DGLAP
equation from

$$G(x,Q^2)=G(x_1,Q^2_1)+\Delta G(x,Q^2)$$
$$=G(x_1,Q^2_1)-2\int^{Q^2}_{Q^2_{1min}}\frac{dQ^2_1}{Q^4_1}\int_{x}^{1/2}\frac{dx_1}{x_1}\frac{x}{x_1}
{\cal K}_{MD-DGLAP}\left(\frac{x}{x_1},\alpha_s\right)
G^{(2)}(x_1,Q_1^2)$$
$$+\int^{Q^2}_{Q^2_{1min}}\frac{dQ^2_1}{Q^4_1}\int_{x/2}^{1/2}\frac{dx_1}{x_1}\frac{x}{x_1} {\cal
K}_{MD-DGLAP}\left(\frac{x}{x_1},\alpha_s\right)
G^{(2)}(x_1,Q_1^2),\eqno(11)$$ where a power suppression factor
$1/Q^2_1$ has been extracted from the evolution kernel. The
positive and negative terms are from the contributions of the
diagrams Fig.2(a) and (b), respectively. The TOPT calculations [8]
give

$${\cal K}^{MD-DGLAP}\frac{dx_1}{x_1}$$
$$=\frac{\alpha^2_s}{8} \frac{N^2_c}{N^2_c-1}\frac{(2x_1-x)(72x_1^4-48x_1^3x+140x_1^2x^2-116x_1x^3+29x^4)}{x_1^5x}dx_1$$
$$\stackrel{x\ll x_1}{\longrightarrow}18\alpha^2_s\frac{N^2_c}{N^2_c-1}\frac{dx_1}{x}. \eqno(12)$$

    The gluon correlation function $G^{(2)}$ is a generalization of the
gluon distribution beyond the leading twist. It is usually
modelled as the product of two gluon distributions [6]. For
example,

$$G^{(2)}(x,Q^2)=R_GG^2(x,Q^2),\eqno(13)$$ where $R_G=1/(\pi R^2)$ is a correlation coefficient with the dimension
$[L^{-2}]$, $R$ is the effective correlation length of two
recombination gluons and we take $R=5 GeV^{-1}$  (the~ radius~ of~
a ~nucleon). Setting Eqs.(12) and (13) to Eq.(11), we obtain the
modified DGLAP equation combining DGLAP dynamics at small $x$

$$\frac{\partial G(x,Q^2)}{\partial\ln Q^2}$$
$$=\frac{\alpha_sN_c}{\pi}\int^1_x
\frac{dx_1}{x_1}G(x_1,Q^2)-\frac{36\alpha_s^2}{\pi
Q^2R^2}\frac{N^2_c}{N^2_c-1}
\int_x^{1/2}\frac{dx_1}{x_1}G^2(x_1,Q^2)
$$
$$+\frac{18\alpha_s^2}{\pi
Q^2R^2}\frac{N_c^2}{N_c^2-1}
\int_{x/2}^{1/2}\frac{dx_1}{x_1}G^2(x_1,Q^2)$$
$$=\frac{\alpha_sN}{\pi}\int^1_x
\frac{dx_1}{x_1}G(x_1,Q^2)-\frac{18\alpha_s^2}{\pi
Q^2R^2}\frac{N^2_c}{N_c^2-1}
\int_{x}^{1/2}\frac{dx_1}{x_1}G^2(x_1,Q^2)$$
$$+\frac{18\alpha_s^2}{\pi Q^2R^2}
\frac{N^2_c}{N^2_c-1}\int_{x/2}^{x}\frac{dx_1}{x_1}G^2(x_1,Q^2) .
\eqno(14)$$

    It is interesting to compare this small-$x$ version of the modified DGLAP
equation with the GLR-MQ equation, which is [4]

$$\frac{\partial G(x,Q^2)}{\partial\ln Q^2}$$
$$=\frac{\alpha_sN_c}{\pi}\int^1_x
\frac{dx_1}{x_1}G(x_1,Q^2)-\frac{36\alpha_s^2}{8
Q^2R^2}\frac{N^2_c}{N^2_c-1}
\int_x^{1/2}\frac{dx_1}{x_1}G^2(x_1,Q^2),  \eqno(15)$$ where

$$G^{(2)}(x,Q^2)=\frac{9}{8\pi R^2}G^2(x,Q^2),\eqno(16)$$ is assumed.
Although both of the two equations have been hoped to describe the
corrections of the gluon recombination to the linear DGLAP
equation at the $DLLA$, they have different forms due to the
following reasons:

    (1) The GLR-MQ equation is derived basing on the following two works. One is the idea
that the negative shadowing effects arise from the interference
processes of the gluon recombination. Gribov, Levin and Ryskin
used the AGK cutting rules to show that the contributions from the
real cut diagrams (see Fig.2(a)) and the interferant cut diagrams
(see Fig.2(b)) only differ in the numerical weights, which is 2
and -4, respectively. Thus, the net effects of the gluon
recombination can be simply calculated by multiplying the real
diagrams by a negative weight. Obviously, the antishadowing
effects in this approach are completely cancelled by the shadowing
effects and the resulting evolution equation violates the momentum
conservation. Another work is the calculation of the recombination
functions at the $DLLA$ in a covariant perturbation theory by
Mueller and Qiu [4], where a special treatment was used to remove
the infrared (IR) singularities in the gluonic twist-4 coefficient
functions. The gluon recombination functions in the GLR-MQ
equation are separated out from the above mentioned IR-safe
coefficient functions.

    (2) In the derivation of the modified DGLAP equation, the TOPT was first used to
establish the relations among all the relevant cut diagrams in
Fig.2 [6].  We showed that the shadowing and antishadowing effects
share the same recombination function but occupy different
kinematic regions. On the other hand, the contributions of the two
virtual diagrams Fig.2(c) and (d) cancel against each other. Thus,
the net effects depend not only on the local value of the gluon
distribution at the observed point, but also on the shape of the
gluon distribution when the Bjorken variable goes from $x$ to
$x/2$. In consequence, the shadowing effects in the evolution
process will be obviously weakened by the antishadowing effects if
the distribution is steeper [11].  We found that the same TOPT
framework allows us to calculate the recombination functions
including quarks and gluons in the whole region of $x$ at the
$LL(Q^2)A$.  Consequently, the momentum conservation is restored
in the modified DGLAP equation [6]. We also showed in [8] that the
gluon recombination functions can be reasonably separated from the
twist-4 coefficient functions in the TOPT approach without the
above mentioned special treatment, though the results of  the two
methods only differ by a constant at the $DLLA$.

    In next step, we consider the recombination of two unintegrated
gluon distribution functions, which obey the BFKL equation. The
cut diagram corresponding to Fig.1(a) is Fig.1(b), where the
dashed box implies a new recombination kernel of two BFKL
solutions. The initial gluons couple with the gluonic dipole and
it means that two initial legs evolve according to the BFKL
dynamics. One can expect that ${\cal K}_{MD-BFKL}$ is more
complicated than ${\cal K}_{MD-DGLAP}$. As an approximate model,
we use kernel ${\cal K}_{MD-DGLAP}$ to replace the kernel ${\cal
K}_{MD-BFKL}$ in Fig.1(b). In the concrete, the contributions of
two correlated unintegrated distribution functions $F^{(2)}$ to
the measured (integrated) distribution $G$ via the recombination
processes are

$$\Delta G(x,Q^2)=
-2\int^{Q^2}_{Q^2_{min}}\frac{d\underline{k}^2}{\underline{k}^4}\int_{x}
^{1/2}\frac{dx_1}{x_1}\frac{x}{x_1} {\cal
K}_{MD-DGLAP}\left(\frac{x}{x_1},\alpha_s\right)
F^{(2)}(x_1,\underline{k}^2) $$
$$+\int^{Q^2}_{Q^2_{min}}\frac{d\underline{k}^2}{\underline{k}^4}\int_{x/2}
^{1/2}\frac{dx_1}{x_1}\frac{x}{x_1} {\cal
K}_{MD-DGLAP}\left(\frac{x}{x_1},\alpha_s\right)
F^{(2)}(x_1,\underline{k}^2). \eqno(17)$$ Thus, we have

$$\Delta F(x,\underline{k}^2)=
\left.Q^2\frac{\partial\Delta G(x,Q^2)} {\partial Q^2}
\right\vert_{Q^2=\underline{k}^2}$$
$$=-\frac{2}{\underline{k}^2}\int_{x}
^{1/2}\frac{dx_1}{x_1}\frac{x}{x_1} {\cal
K}_{MD-DGLAP}\left(\frac{x}{x_1},\alpha_s\right)
F^{(2)}(x_1,\underline{k}^2)$$
$$+\frac{1}{\underline{k}^2}\int_{x/2}
^{1/2}\frac{dx_1}{x_1}\frac{x}{x_1} {\cal
K}_{MD-DGLAP}\left(\frac{x}{x_1},\alpha_s\right)
F^{(2)}(x_1,\underline{k}^2). \eqno(18)$$  Using Eqs.(6) and (12)
we obtain the corrections to the evolution of the unintegrated
gluon distribution along small $x$ direction

$$-x\frac{\partial F(x,\underline{k}^2)}{\partial x}
=-36\alpha_s^2\frac{N_c^2}{N_c^2-1}\frac{1}{\underline{k}^2}
F^{(2)}(x,\underline{k}^2)+18\alpha_s^2\frac{N^2_c}{N^2_c-1}\frac{1}{\underline{k}^2}
F^{(2)}(\frac{x}{2},\underline{k}^2) . \eqno(19)$$ Combining with
the BFKL equation, we obtain a unitarized BFKL equation

$$-x\frac{\partial F(x,\underline{k}^2)}{\partial x}$$
$$=\frac{\alpha_{s}N_c\underline{k}^2}{\pi}\int_{\underline{k}'^2_{min}}^{\infty} \frac{d \underline{k
}'^2}{\underline{k}'^2}\left\{\frac{F(x,\underline{k}'^2)-F(x,\underline{k}^2)}
{\vert
\underline{k}'^2-\underline{k}^2\vert}+\frac{F(x,\underline{k}^2)}{\sqrt{\underline{k}^4+4\underline{k}'^4}}\right\}$$
$$-\frac{36\alpha_s^2}{\pi\underline{k}^2R^2}\frac{N_c^2}{N_c^2-1}
F^2(x,\underline{k}^2)+\frac{18\alpha_s^2}{\pi\underline{k}^2R^2}\frac{N_c^2}{N_c^2-1}
F^2\left(\frac{x}{2},\underline{k}^2\right), \eqno(20)$$ where
similar to Eq. (13) we define

$$F^{(2)}(x,\underline{k}^2)=\frac{1}{\pi
R^2}F^2(x,\underline{k}^2).\eqno(21)$$  Eq.(20) is our unitarized
BFKL equation for the unintegrated gluon distribution. Since the
gluon distribution becomes flatter near the saturation limit, in
this range we can take the approximation

    $$F\left(\frac{x}{2},\underline{k}^2\right)
\simeq F(x,\underline{k}^2), \eqno(22)$$ in Eq.(20), and get

    $$-x\frac{\partial F(x,\underline{k}^2)}{\partial x}$$
$$=\frac{\alpha_{s}N_c\underline{k}^2}{\pi}\int_{\underline{k}'^2_{min}}^{\infty} \frac{d \underline{k
}'^2}{\underline{k}'^2}\left\{\frac{F(x,\underline{k}'^2)-F(x,\underline{k}^2)}
{\vert
\underline{k}'^2-\underline{k}^2\vert}+\frac{F(x,\underline{k}^2)}{\sqrt{\underline{k}^4+4\underline{k}'^4}}\right\}$$
$$-\frac{18\alpha_s^2}{\pi\underline{k}^2R^2}\frac{N_c^2}{N_c^2-1}
F^2(x,\underline{k}^2). \eqno(23)$$

\newpage
\begin{center}
\section{Numerical analysis}
\end{center}

        We make some numerical calculations to illustrate the antishadowing effects.
For simplicity, we fix the coupling constant to be $\alpha_s=0.3$.
Eq.(20) is the corrections of the gluon recombination to the BFKL
evolution equation. Therefore, we use a parameter form of the BFKL
solution as the input distribution at $x_0=10^{-2}$ [12]

$$
F(x_0,\underline{k}^2)=\beta\sqrt{\underline{k}^2}\frac{x_0^{-\lambda_{BFKL}}}{\sqrt
{\ln{\frac{1}{x_0}}}}exp\left(-\frac{\ln^2(\underline{k}^2/\underline{k}^2_s)}{2\lambda''\ln
(1/x_0)}\right), \eqno(24)
$$ where $\lambda_{BFKL}=12\alpha_s/(\pi \ln 2)$ and $\lambda''\simeq
32.1 \alpha_s$. The parameter $\underline{k}^2_s=1 GeV^2$ is of
nonperturbative origin and the normalization constant $\beta\sim
0.01$. On the other hand, we take $\underline{k}'^2_{min}=0.01
GeV^2$ in Eq.(20).

     We assume that Eq.(20) begins to work from the BFKL-region.
In this case, the distribution $F(x_0,\underline{k}^2)$ has a
steeper $x$-dependence at the starting point $x_0$ of the
evolution where the shadowing and antishadowing effects almost
cancel out. Thus, we compute the value of
$F(x_i/2,\underline{k}^2)$ at the i-th step of the evolution from
$F(x_i,\underline{k}^2)$ using the BFKL equation, i.e.,

$$F(x_i,\underline{k}^2)
\stackrel{BFKL-path}{\longrightarrow}F\left(\frac{x_i}{2},\underline{k}^2\right)\equiv
F_{BFKL}\left(\frac{x_i}{2},\underline{k}^2\right) .\eqno(25)$$
The numerical results show a faster divergence of the distribution
$F(x,\underline{k}^2)$ due to the antishadowing effects that
cancel out or even outweigh the shadowing effects in the typical
BFKL-solution. Of course, we can not expect such infinite growth
of the gluons at small $x$. This unphysical result appears because
the BFKL dynamics are not constrained by the energy-momentum
conservation, which must be added from the outside of the
evolution equation. According to the experiences in the solutions
of the modified DGLAP equation, the net antishadowing effects in a
small $x$ region always accompany the stronger net shadowing
effects in the smaller $x$ region [11]. One can expect that the
evolution asymptotically approaches the BK dynamics under the
action of the net shadowing effects. Therefore, we modify the
program (25) to

$$F\left(\frac{x_i}{2},\underline{k}^2\right)=F_{Shadowing}\left(\frac{x_i}{2},\underline{k}^2\right)
+\frac{F_{BFKL}\left(\frac{x_i}{2},\underline{k}^2\right)-F_{Shadowing}\left(\frac{x_i}{2},\underline{k}^2\right)}{i\eta^F
-\eta^F+1} ,\eqno(26)$$ where
$F_{Shadowing}(x_i/2,\underline{k}^2)$ indicates that the
evolution from $x_i$ to $x_i/2$ is controlled by Eq.(23). The
parameter $\eta^F$ implies the different velocities approaching
the BK dynamics. There is a limit value of $\eta^F_{min}\simeq
0.001$ and the solution of Eq.(20) becomes divergent when
$\eta^F<\eta^F_{min}$.  As an example, we take
$\eta^F=\eta^F_{min}$ and divide the evolution into the
$100$-steps on $\vert\Delta\log x\vert=1$.

    In Fig.3(a) we plot the unintegrated gluon distribution
function (solid curves) as a function of $x$ for two different
values of $\underline{k}^2$ and where $\eta^F_{min}=0.001$. The
possible solutions of Eq.(20) should lie between the solid and
point curves since $\eta^F>\eta^F_{min}$. For comparison, the
results of two calculations based on the BFKL equation (dashed
curves) and Eq.(23) without the antishadowing effects (point
curves) are listed. One can find that Eq.(20) keeps the
BFKL-behavior in a larger evolution region than Eq.(23) due to the
antishadowing effects.

    The $\underline{k}^2$-dependence of the unintegrated gluon distribution
function with $\eta^F_{min}=0.001$ is given in Fig.3(b). The
saturation of distribution is usually defined as a limit form,
which is insensitive to $x$ or $\underline{k}^2$. We have not
observed such saturation phenomenon in a broad kinematical range,
although the obvious suppressions of the net shadowing effects can
be found in Fig.3. The reason is that the factor
$1/\underline{k}^2$ in both the Eqs.(20) and (23) suppresses the
contributions of the nonlinear terms at larger values of
$\underline{k}^2$.

   We give the integrated gluon density using Eq.(3) in Fig.4.
We still have not observed the saturation limit on the
$Q^2$-dependence of the distribution $G(x,Q^2)$ in Fig.4(b). It is
interesting that this conclusion is qualitatively consistent with
our previous work [11], where the corrections of the same gluon
recombination kernel to the DGLAP evolution equation are
considered.  Of course, to obtain more practical predictions about
the gluon distribution at small $x$, further corrections should be
considered, for example, the corrections from the running
$\alpha_s$, the BFKL dynamics in the IR-region
$\underline{k}'^2\ll 1 GeV^2$, and in particular, the choice of
the input distribution. However, the important distinction between
the antishadowing effects and shadowing effects as demonstrated in
Fig.4 should remain after those corrections are incorporated.

\newpage
\begin{center}
\section{Discussions}
\end{center}

    We try to compare the nonlinear evolution equation (20) with the
BK equation, which is originally written in the transverse
coordinate space for the scattering amplitude.  There are
different definitions of the scattering amplitude in the QCD
evolution equation.  In this work, we note that the BFKL equation
for the scattering amplitude $N(\underline{k}^2,x)$ has the
following form [13],

$$-x\frac{\partial N(\underline{k}^2,x)}{\partial x}$$
$$=\frac{\alpha_{s}N_c}{\pi}\int_{\underline{k}'^2_{min}}^{\infty} \frac{d \underline{k
}'^2}{\underline{k}'^2}\left\{\frac{\underline{k}'^2N(\underline{k}'^2,x)-\underline{k}^2N(\underline{k}^2,x)}
{\vert
\underline{k}'^2-\underline{k}^2\vert}+\frac{\underline{k}^2N(\underline{k}^2,x)}{\sqrt{\underline{k}^4+4\underline{k}'^4}}\right\}
, \eqno(27)$$ one can find that $F(x,\underline{k}^2)$ and
$N(\underline{k}^2,x)$ differ by a simple scale translation
$F(x,\underline{k}^2)\sim \underline{k}^2N(\underline{k}^2,x)$.
Thus, we define the scattering amplitude

$$N(\underline{k}^2,x)\equiv
\frac{27\alpha_s}{4\underline{k}^2R^2}F(x,\underline{k}^2)
,\eqno(28)$$ where we choose such constants on the right-hand side
of equation that the following resulting equation is consistent
with the BK equation at the saturation limit.

    Using Eq.(28) we rewrite Eq.(20) as

$$-x\frac{\partial N(\underline{k}^2,x)}{\partial x}$$
$$=\frac{\alpha_{s}N_c}{\pi}\int_{\underline{k}'^2_{min}}^{\infty} \frac{d \underline{k
}'^2}{\underline{k}'^2}\left\{\frac{\underline{k}'^2N(\underline{k}'^2,x)-\underline{k}^2N(\underline{k}^2,x)}
{\vert
\underline{k}'^2-\underline{k}^2\vert}+\frac{\underline{k}^2N(\underline{k}^2,x)}{\sqrt{\underline{k}^4+4\underline{k}'^4}}\right\}$$
$$-2\frac{\alpha_sN_c}{\pi}N^2(\underline{k}^2,x)+\frac{\alpha_sN_c}{\pi}N^2\left(\underline{k}^2,\frac{x}{2}\right). \eqno(29)$$
Interestingly, this equation is consistent with the BK equation (in
the impact parameter-independent case) in momentum space near the
saturation limit due to

$$N\left(\underline{k}^2,\frac{x}{2}\right)\simeq N(\underline{k}^2,x). \eqno(30)$$

For the cylindrically symmetric solution, we use the following
transformation according to [14]

$$N(k,x)=\int
\frac{d^2\underline{r}}{2\pi}\exp(-i\underline{k}\cdot\underline{r})
\frac{N(r,x)}{r^2}$$
$$=\int^{\infty}_{0}\frac{dr}{r}J_0(kr)N(r,x),\eqno(31)$$ where $J_0$
is the Bessel function, one can find that Eq.(29) has the
following form in transverse coordinate space

$$-x\frac{\partial N(r_{b0},x)}{\partial x}$$
$$=\frac{\alpha_{s}N_c}{2\pi^2}\int d^2 \underline{r}_{c}
\frac{\underline{r}_{b0}^2}{\underline{r}_{bc}^2\underline{r}_{c0}^2}\left[N(r_{bc},x)+N(r_{c0},x)-N(r_{b0},x)\right.$$
$$-2N(r_{bc},x)N(r_{c0},x)+N\left(r_{bc},\frac{x}{2}\right)N\left(r_{c0},\frac{x}{2}\right)],\eqno(32)$$
which reduces to the BK equation (in the impact
parameter-independent case) at the saturation limit

$$-x\frac{\partial N(r_{b0},x)}{\partial x}$$
$$=\frac{\alpha_{s}N_c}{2\pi^2}\int d^2 \underline{r}_c
\frac{\underline{r}^2_{b0}}{\underline{r}_{bc}^2\underline{x}_{c0}^2}\left[N(r_{bc},x)+N(r_{c0},x)-N(r_{b0},x)
-N(r_{bc},x)N(r_{c0},x)\right]. \eqno(33)$$

     We take the Golec-Biernat-W$\ddot{u}$sthoff model [15] as the input amplitude,
i.e.,

$$N(r, x_0)=1-\exp\left[-\frac{r^2Q'^2_s}{4}\right], \eqno(34)$$
where $Q'_s\simeq~1.4GeV$ relates to the saturation scale. The
computing program of Eq.(32) is

$$N\left(\underline{k}^2,\frac{x_i}{2}\right)=N_{BK}\left(\underline{k}^2,\frac{x_i}{2}\right)
+\frac{N_{BFKL}\left(\underline{k}^2,\frac{x_i}{2}\right)-N_{BK}\left(\underline{k}^2,\frac{x_i}{2},\right)}{i\eta^N
-\eta^N+1}, \eqno(35)$$ where $N_{BK}(\underline{k}^2,x_i/2)$
indicates that the evolution from $x_i$ to $x_i/2$ is according to
Eq.(33).  The minimum parameter $\eta^N_{min}\simeq 0.01$, which
is larger than $\eta^F_{min}$ since the nonlinear effects in
Eq.(32) are stronger than that in Eq.(20).

    Fig.5 shows the solutions (solid curves) of Eq.(32) with $\eta^N_{min}\simeq
0.01$.  For comparison, we give the solutions of the BK equation
(point curves) and the linear part of Eq.(32) (dashed curves). The
possible solutions of Eq.(32) should lie between the solid and
point curves since $\eta^N>\eta^N_{min}$. Fig.6 is such a possible
solution with $\eta_N=0.1$.  One can find that the antishadowing
effects lead to quite different behaviors between the amplitude
$N(r,x)$ and the distribution $F(x,\underline{k}^2)$. The
antishadowing effects on the scattering amplitude sharpen the
transition form of the amplitude $N(r,x)$ and even violate the
unitarity due to $N> 1$ if we take the definition Eq.(28). The
reasons for the stronger antishadowing effects are that (1) we use
a normalized definition about the amplitude in Eq.(28) so that the
coefficients of the nonlinear terms are the same magnitude as that
of the linear terms in Eq.(32); (2) the suppression factor
$1/\underline{k}^2$ is absorbed by the scattering amplitude due to
Eq.(28).

       It is interesting that the momentum compensation for the
shadowing effects was also discussed using a modified nonlinear
evolution equation in Ref.[16]. In the impact
parameter-independent case this equation reads

$$-x\frac{\partial N(r_{b0},x)}{\partial x}=\frac{\alpha_{s}N_c}{2\pi^2}\int d^2 \underline{r}_{c}
\frac{\underline{r}_{b0}^2}{\underline{r}_{bc}^2\underline{r}_{c0}^2}$$
$$\left\{2N(r_{bc},x)-N(r_{bc},x)N(r_{c0},x)
-\frac{\partial}{\partial
Y}[2N(r_{bc},x)-N(r_{bc},x)N(r_{c0},x)]\right\},\eqno(36)$$ where
$Y=\ln(1/x)$ and the last term is the leading order DGLAP
corrections.  For comparison, we rewrite our Eq.(32) as

$$-x\frac{\partial N(r_{b0},x)}{\partial x}$$
$$\simeq \frac{\alpha_{s}N_c}{2\pi^2}\int d^2
\underline{r}_{c}
\frac{\underline{r}_{b0}^2}{\underline{r}_{bc}^2\underline{r}_{c0}^2}
\left\{N(r_{bc},x)+N(r_{c0},x)-N(r_{b0},x)
-2N(r_{bc},x)N(r_{c0},x)\right.$$
$$\left.+N(r_{bc},x)N(r_{c0},x)
-\frac{x}{2}\frac{\partial}{\partial
x}[N(r_{bc},x)N(r_{c0},x)]\right\}$$
$$=\frac{\alpha_{s}N_c}{2\pi^2}\int d^2 \underline{r}_{c}
\frac{\underline{r}_{b0}^2}{\underline{r}_{bc}^2\underline{r}_{c0}^2}\left\{N(r_{bc},x)+N(r_{c0},x)-N(r_{b0},x)
-N(r_{bc},x)N(r_{c0},x)\right.$$
$$\left.+\frac{1}{2}\frac{\partial}{\partial
Y}[N(r_{bc},x)N(r_{c0},x)]\right\}. \eqno(37)$$ Although Eqs.(36)
and (37) have similar nonlinear antishadowing terms (only differ
by a factor $1/2$), one can find that the main difference between
these two equations is that the same momentum compensation
mechanism acts on both linear and nonlinear terms in Eq.(36),
while the corresponding corrections in Eq.(37) are from the
virtual and recombination processes, respectively.

     Although we have discussed the contributions of the
antishadowing effects to the unitarized BFKL equation, we note,
however, that the following problems remain to be solved: (i) As
we have emphasized, the replacement of the recombination functions
for two BFKL-solutions with ${\cal K}_{MD-DGLAP}$ in this work is
an approximate method. The unitarized BFKL equation near the
saturation range should take a new form ; (ii) How to determine
the parameter $\eta^F$ in Eq.(26)? Perhaps, the most reliable test
of the antishadowing effects would be a comparison to the nuclear
antishadowing effects, which has been observed in the EMC effect
[17]. However, the kinematic region of the nuclear antishadowing
effects is about $0.05<x<0.3$, where we should combine the DGLAP
dynamics in Eq.(20) and it is beyond the scope of the present
paper; (iii) The QCD evolution should be performed with running
$\alpha_s$ in the numerical procedure. These problems will be our
following subjects.

    In conclusion, we presented the corrections of the gluon recombination to the
BFKL equation and they lead to a new unitarized nonlinear
evolution equation, which incorporates both shadowing and
antishadowing effects. The new equation reduces to the BK equation
near the saturation limit. The numerical solutions of the equation
show that the antishadowing effects have a sizeable influence on
the gluon distribution function in the preasymptotic regime.

\vspace{0.3cm}

\noindent {\bf Acknowledgments}: This work was supported by
National Natural Science Foundations of China 10135060, 10475028
and 10205004.

\newpage

Figure Captions

Fig. 1 The cut diagram containing $4\rightarrow 2$ gluon
recombination kernel (the dashed box) in one step QCD evolution:
(a) for the modified DGLAP equation and (b) for the modified BFKL
equation, where two legs before fusion are resummed using the
DGLAP and BFKL equations in (a) and (b), respectively. The dark
area implies the probe.

Fig. 2 The cut diagrams of the gluon recombination kernels for (a)
the real processes that yield the antishadowing effects; (b) the
interference processes that yield the shadowing effects in the
modified DGLAP equation, (c) and (d) are the corresponding virtual
diagrams.

Fig. 3 (a) The unintegrated gluon distribution function
$F(x,\underline{k}^2)$ as the function of $x$ for different values
of $\underline{k}^2$. The solid-, point- and dashed-curves are the
solutions of Eq.(20) with $\eta^F_{min}=0.001$, Eq.(23) and BFKL
equation, respectively; (b) Similar to (a) but as the function of
$\underline{k}^2$ for different values of $x$.  The possible
solutions of Eq.(20) should lie between the solid and point
curves.

Fig. 4  Similar to Fig.3 but for the gluon distribution function
$G(x,Q^2)$.

Fig. 5 (a) The normalized scattering amplitude $N(r,x)$ as the
function of $x$ for different values of $r$; (b) Similar to (a)
but as the function of $r$ for different values of $x$. The
solid-, point- and dashed-curves are the solutions of Eq.(32) with
$\eta^N_{min}=0.01$, the BK equation (33) and the linear part of
Eq.(32), respectively. The possible solutions of Eq.(32) should
lie between the solid and point curves.

Fig. 6  A possible solution of Eq.(32) with $\eta^N=0.1$ (a) as
the function of $x$, and (b) as the function of $r$.
\newpage
\epsfysize=22cm\epsffile{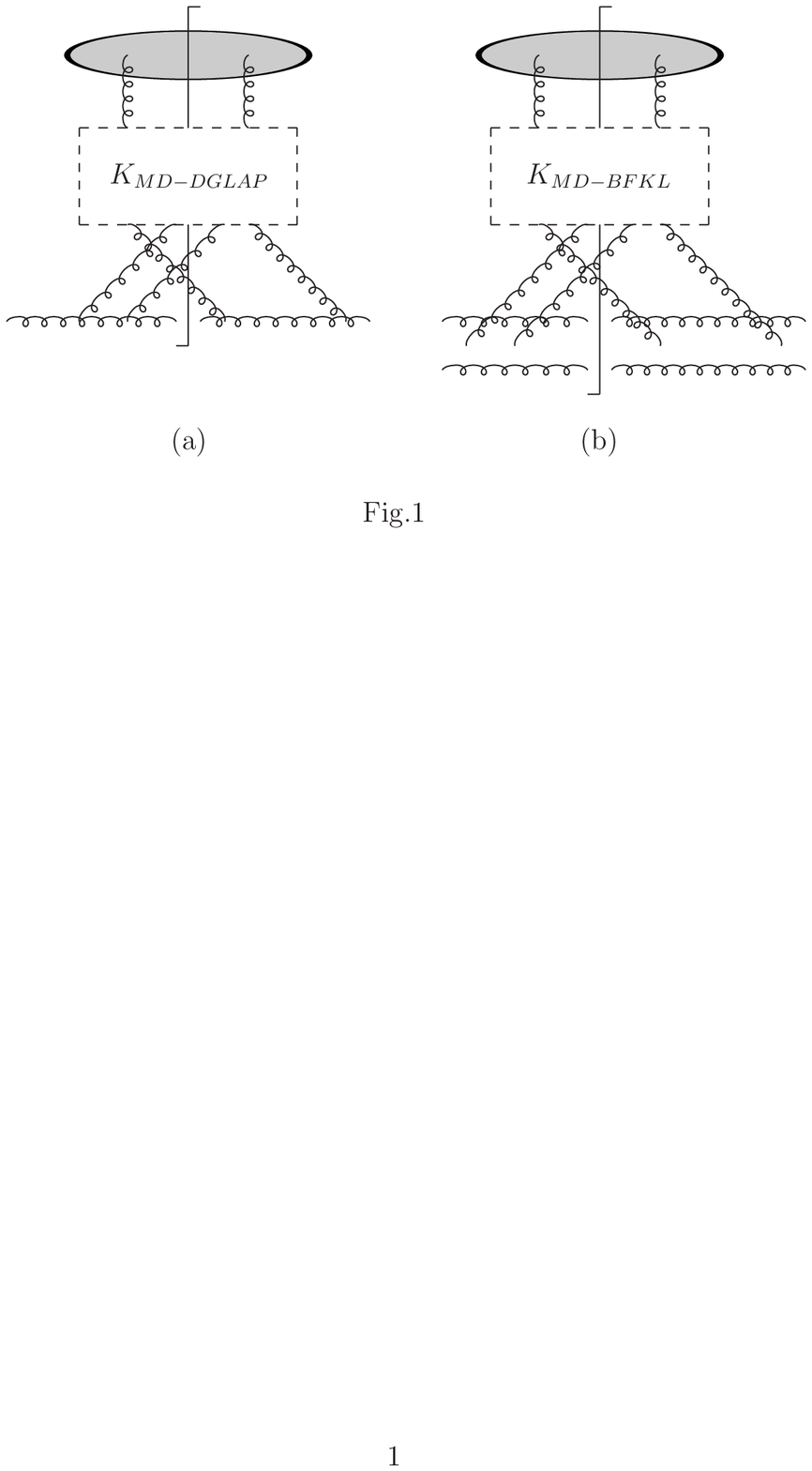}
\newpage
\epsfysize=22cm\epsffile{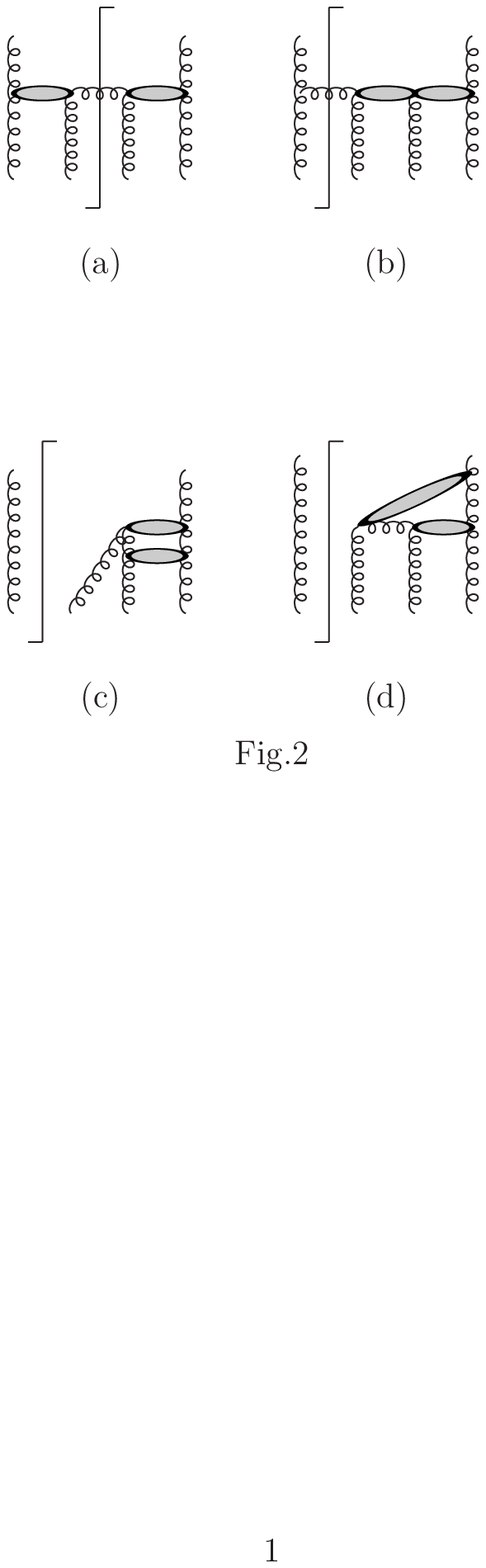}
\newpage
\epsfysize=22cm\epsffile{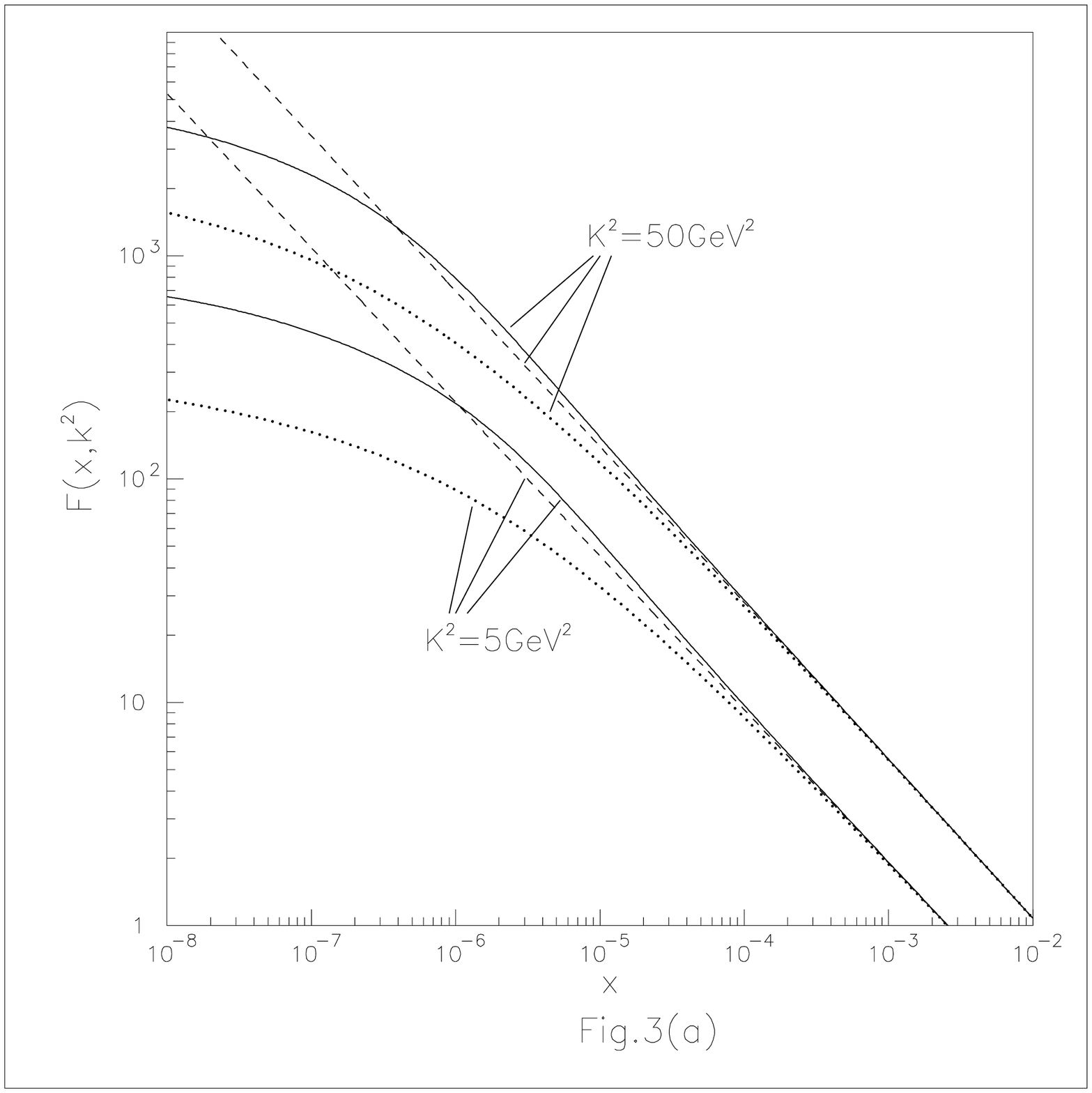}
\newpage
\epsfysize=22cm\epsffile{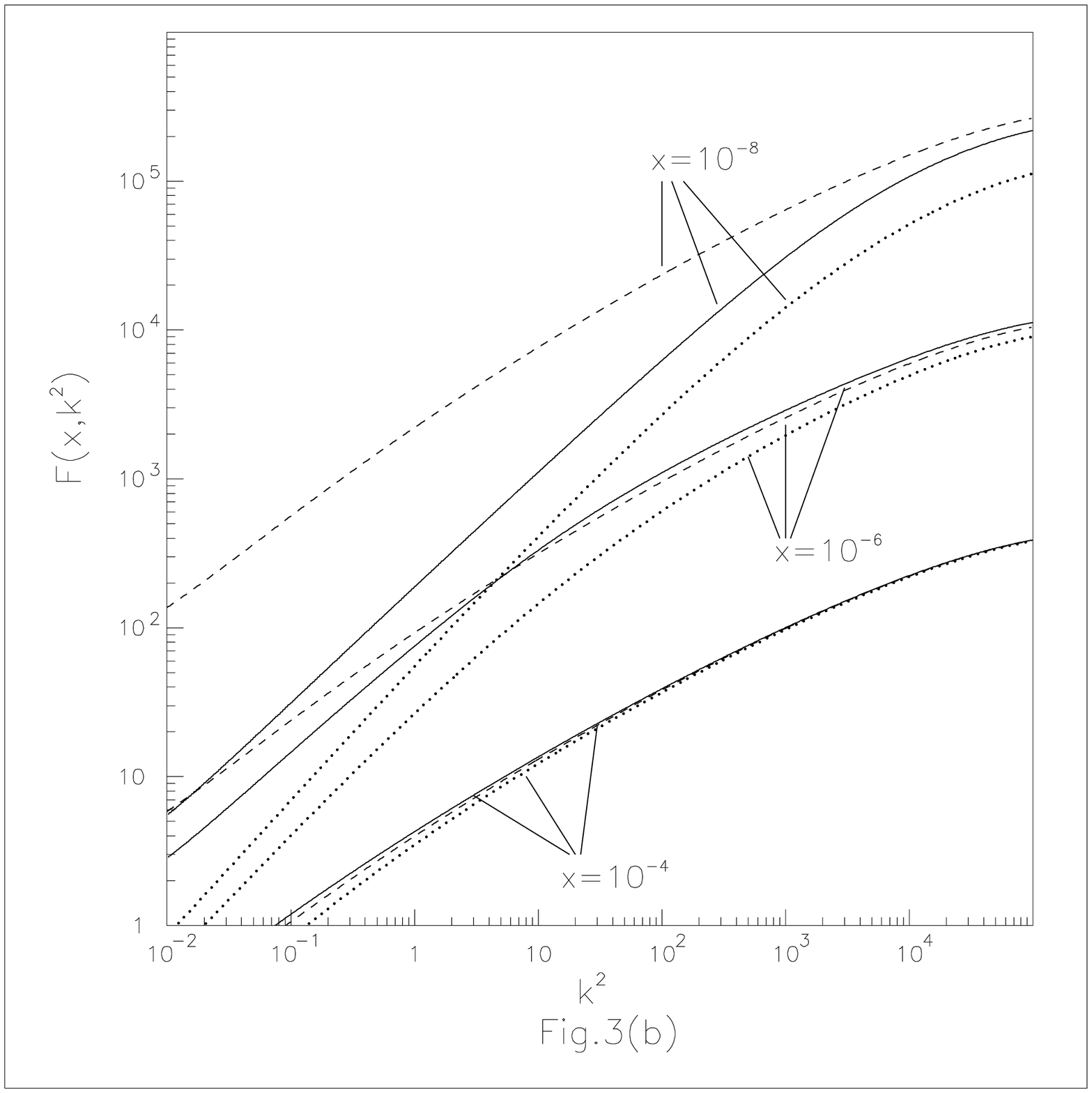}
\newpage
\epsfysize=22cm\epsffile{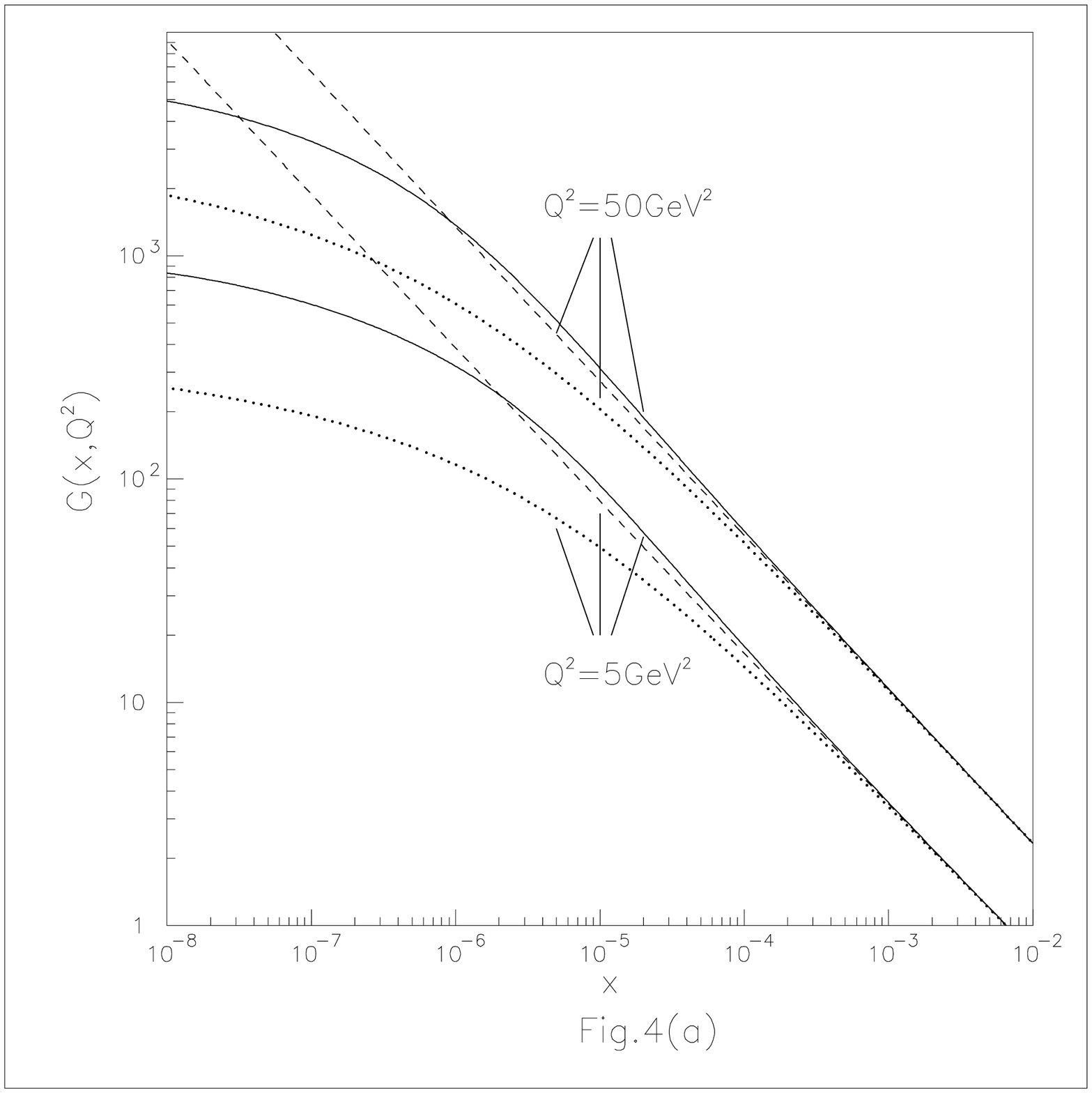}
\newpage
\epsfysize=22cm\epsffile{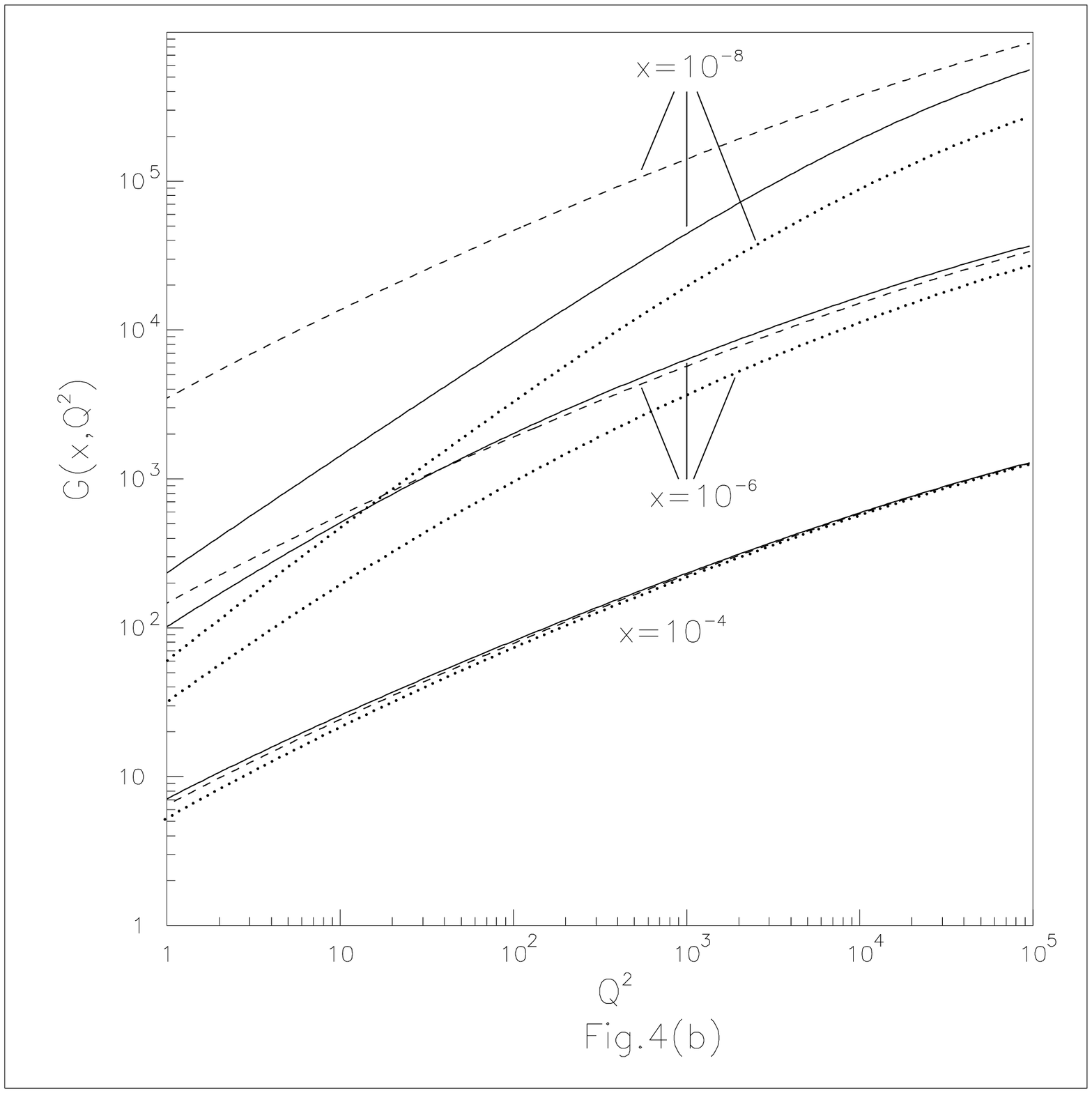}
\newpage
\epsfysize=22cm\epsffile{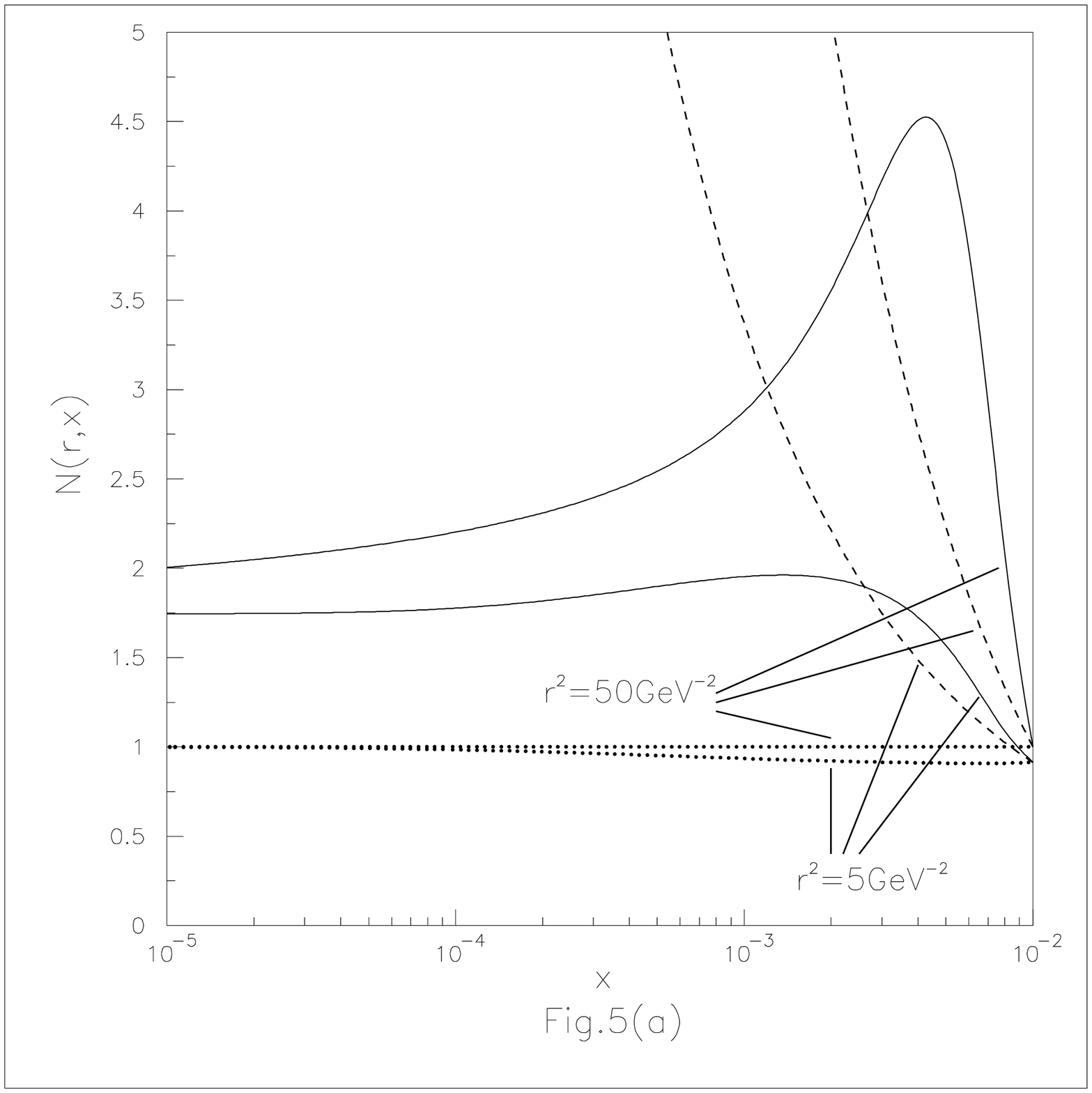}
\newpage
\epsfysize=22cm\epsffile{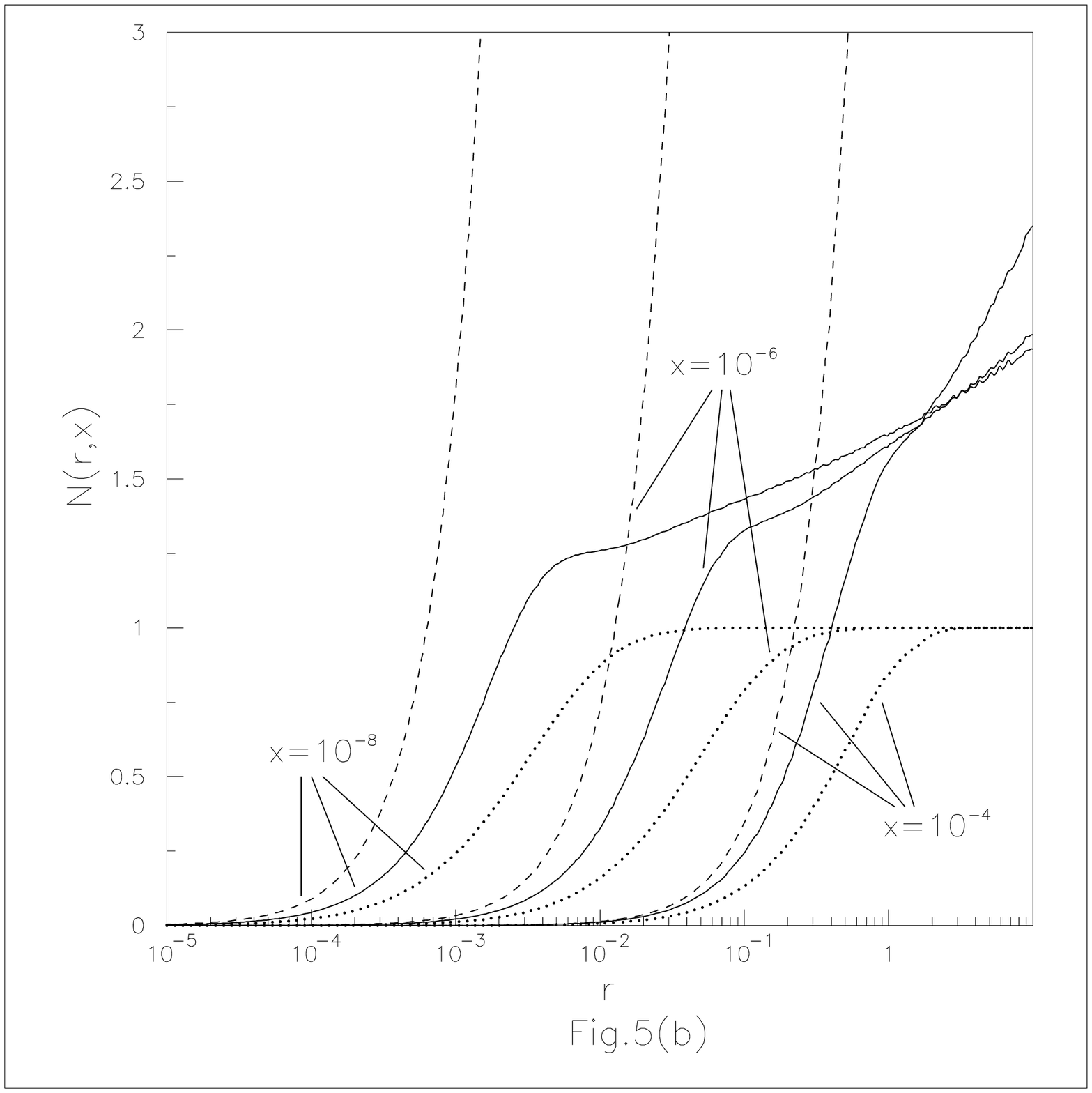}
\newpage
\epsfysize=22cm\epsffile{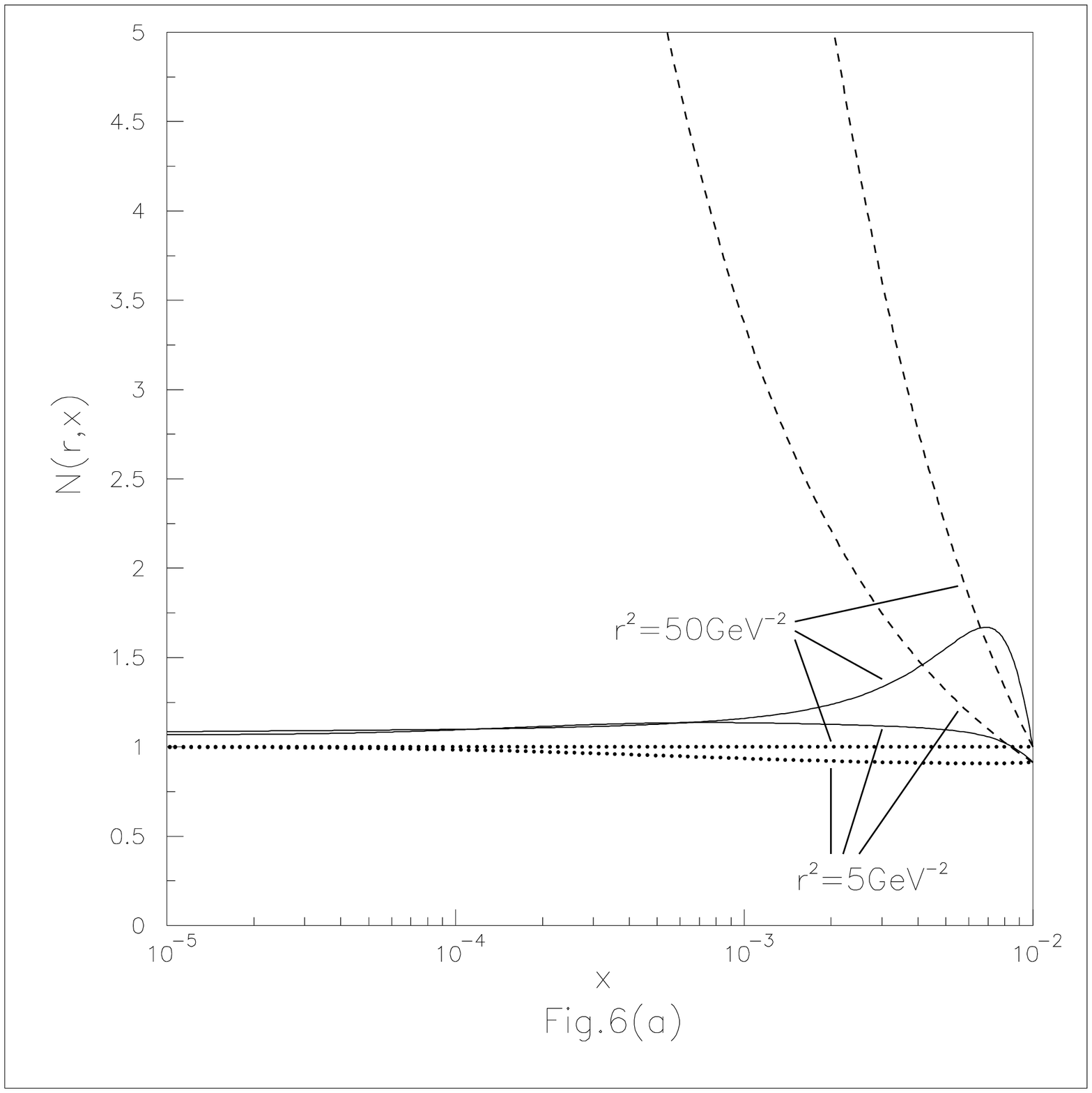}
\newpage
\epsfysize=22cm\epsffile{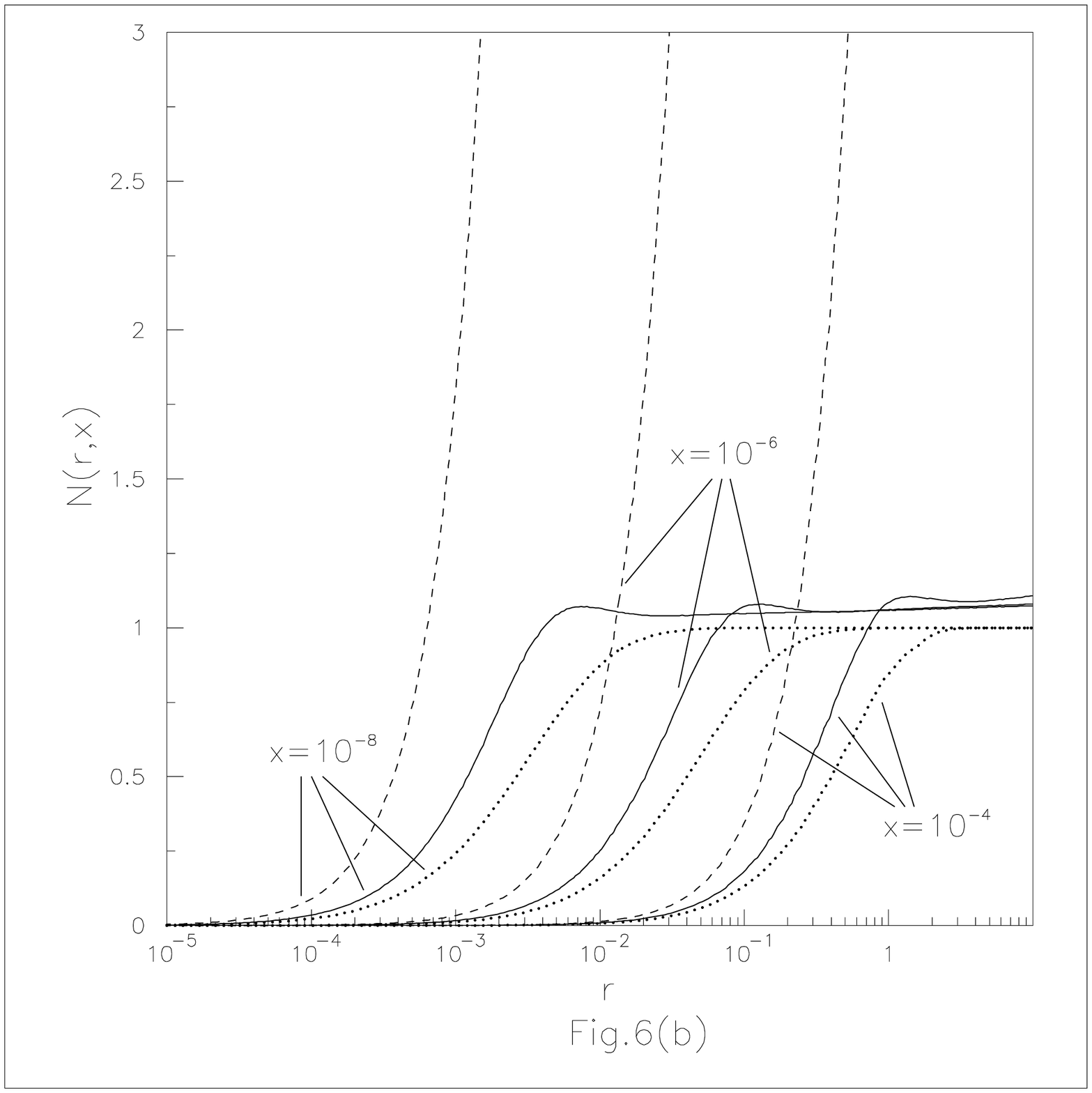}
\end{document}